\documentclass[letterpaper]{article} 
\usepackage[]{aaai23}  
\usepackage{times}  
\usepackage{helvet}  
\usepackage{courier}  
\usepackage[hyphens]{url}  
\usepackage{graphicx} 
\usepackage{natbib}  
\usepackage{caption} 
\usepackage{xspace}
\usepackage{balance}
\usepackage{algorithm}
\usepackage{amsmath}
\usepackage{newfloat}
\usepackage{listings}
\DeclareCaptionStyle{ruled}{labelfont=normalfont,labelsep=colon,strut=off} 
\floatstyle{ruled}
\newfloat{listing}{tb}{lst}{}
\floatname{listing}{Listing}
\title{Feature Decomposition for Reducing Negative Transfer: A Novel Multi-task Learning Method for Recommender System}

\author{
    Jie Zhou,
    Qian Yu\equalcontrib,
    Chuan Luo,
    Jing Zhang
}
\affiliations{
School of Software, Beihang University, China\\
\{zhoujiee, qianyu, chuanluo, zhang\_jing\}@buaa.edu.cn

}

\usepackage{bibentry}

\begin{document}

\maketitle

\begin{abstract}
 In recent years, thanks to the rapid development of deep learning (DL), DL-based multi-task learning (MTL) has made significant progress, and it has been successfully applied to recommendation systems (RS). However, in a recommender system, the correlations among the involved tasks are complex. Therefore, the existing MTL models designed for RS suffer from \textit{negative transfer} to different degrees, which will injure optimization in MTL. We find that the root cause of negative transfer is \textit{feature redundancy} that features learned for different tasks interfere with each other.
To alleviate the issue of negative transfer, we propose a novel multi-task learning method termed Feature Decomposition Network (FDN). The key idea of the proposed FDN is reducing the phenomenon of feature redundancy by explicitly decomposing features into task-specific features and task-shared features with carefully designed constraints. 
We demonstrate the effectiveness of the proposed method on two datasets, a synthetic dataset and a public datasets (i.e., \textit{Ali-CCP}). Experimental results show that our proposed FDN can outperform the state-of-the-art (SOTA) methods by a noticeable margin.
\end{abstract}

\maketitle
\section{Introduction}
\begin{figure}[t]
  \centering
  \includegraphics[width=1.0\linewidth]{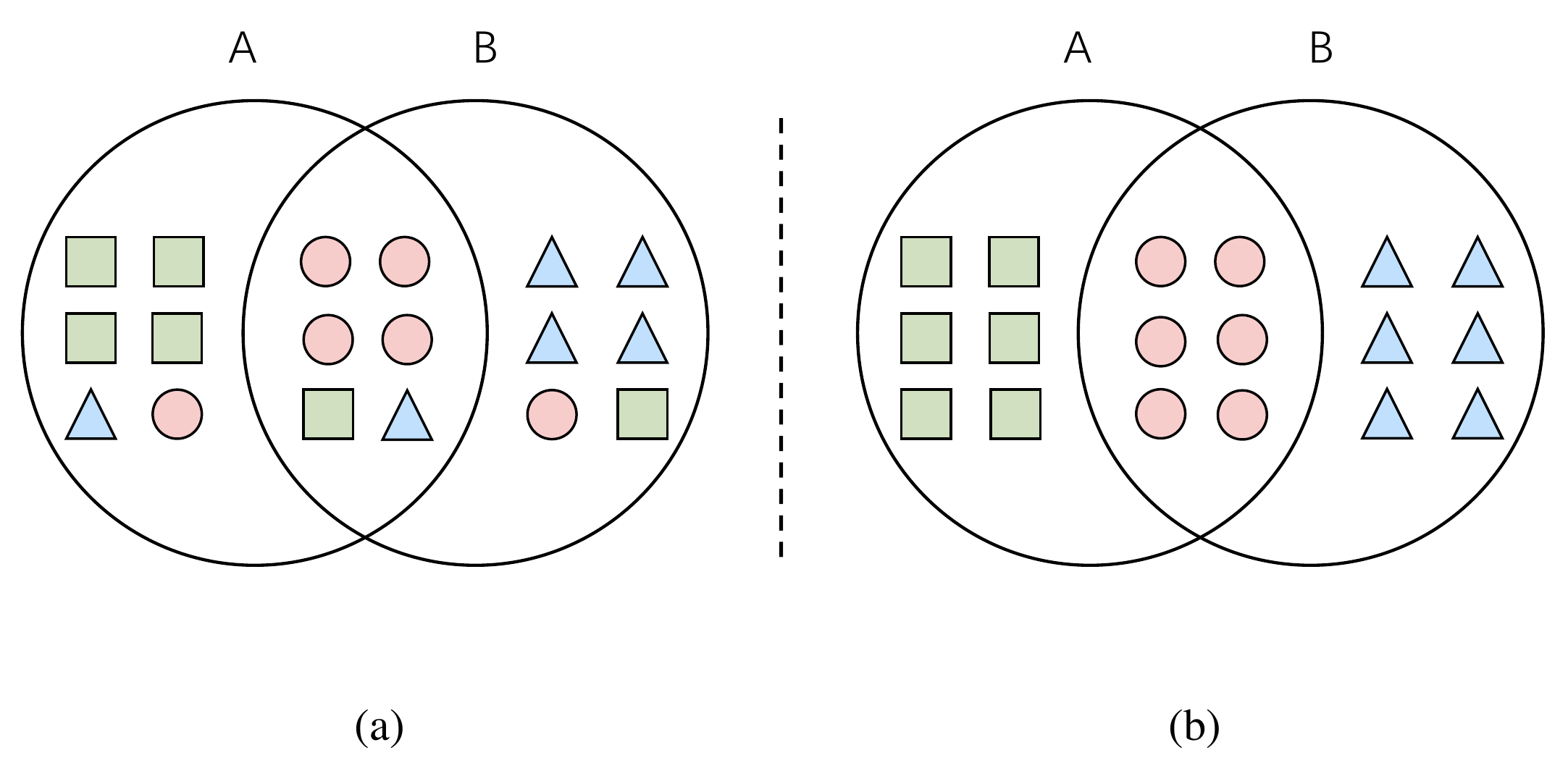}
  \vspace{-0.3cm}
  \caption{The green rectangle indicates the task-specific features of Task A, the blue triangle indicates the task-specific features of Task B, and the pink circle indicates task-shared features.
  }
  \label{fig1}
\end{figure}

The recommendation system (RS) is an effective tool to help users to handle \textit{information overload}. It has been widely used in many commercial fields, such as advertising computing, social networks, and e-commerce\cite{wen2019multi}. 
To improve the user experience, the idea of using multi-task learning (MTL) to satisfy user requirements from multiple aspects has become increasingly popular in the RS community. Many deep learning (DL) based MTL models \cite{ma2018modeling,tang2020progressive} have been proposed for RS. 
However, there is no explicit constraint to force these experts to extract features as required. As a result, the features captured by different experts may still be a mixture of the task-specific feature and the task-shared feature (as shown in Fig.~\ref{fig1}-(a)), which is called \textit{feature redundancy} phenomenon in this paper and will degrade the effectiveness of the model in handling the issue of negative transfer. In the ideal case (as shown in Fig. 1-(b)), feature
spaces should be more pure.

This paper proposes a novel multi-task learning method for recommendation system, named \textbf{F}eature \textbf{D}ecomposition \textbf{N}etwork (FDN). The key idea is to reduce features redundancy by decomposing them with \textit{explicit} constraints. Specifically, we introduce a \textbf{D}e\textbf{C}omposition \textbf{P}air (DCP) to capture task-shared features and task-specific features, respectively. 
We conducted experiments on a public datasets (i.e., Ali-CCP dataset) to demonstrate the effectiveness of our proposed FDN. 


\section{Related works}


In recent years, many MTL models have been proposed for RS. Except for the aforementioned MMoE and PLE, Ma et al. \cite{tang2020progressive} proposed the Entire Space Multi-task Model (ESMM). This model uses features provided by CTR and click- through\&conversion rate (CTCVR) to help build Conversion rate (CVR) model in the whole output space. However, due to the diversity of optimization target types and the number of tasks (generally more than two tasks), ESMM will have the following problems: 1) ESSM cannot easily support other tasks besides classification tasks; 2) With the increase of the number of tasks, ESMM has poor scalability. 

Besides, there have also been some works utilizing AutoML strategy to find a good network structure. Ma et al. \cite{ma2018entire} proposed Sub-Network Routing (SNR) to improve the performance with a flexible parameter sharing strategy by using a routing network. SNR controls sub-networks inputs of each task by binary variables and apply Neural Architecture Search (NAS) strategy to search network structure. Influenced by SNR, Ding et al. \cite{2021MSSM} proposed the multiple level sparse sharing model (MSSM). MSSM proposed field-level and sell-level sparse connection module by binary variables. However, this strategy requires a lot  of parameters, which can be a burden when deploying the model online.
\section{Methodology}

\subsection{Feature redundancy in  MTL}

The negative transfer in MLT is considered to be caused by the fusion of expert's features. Therefore, by explicitly dividing the expert networks into task-shared experts and task-specific experts, PLE \cite{tang2020progressive} model can alleviate the problem of negative transfer. 
However, explicit expert network division strategy can not completely solve the problem. 
As shown in Fig.~\ref{fig1}-(a), due to the lack of explicit constraints on experts, there may be a specific feature space including other spaces' features, that is the so-called feature redundancy. Due to the above phenomena, there will be task-specific features in the task-shared feature space or task-specific features of other tasks in the current task-specific feature space, which will weaken the representation ability of MTL and lead to the problem of negative transfer. In the ideal case (as shown in Fig.~\ref{fig1}-(b)), feature spaces should be more pure. To solve the phenomenon of feature redundancy, we propose DCPs.


\subsection{Preliminaries}

%
Given $K$ tasks (e.g., clicking, sharing, etc), the corresponding labels are $\left\{y_{1}, \ldots, y_{K}\right\}$. $X \in R^{n}$ denotes the input representation where $n$ refers to the dimension of the input. 
The predictions of the $k$-th task can be expressed as $\hat{y}_{k}=f^{MTL}_{k}(X)$,
where $f^{MTL}_{k}(\cdot)$ refers to the network for the $k$-th task of a multi-task learning model.

\begin{figure*}[t]
  \centering
  \includegraphics[width=1.0\linewidth]{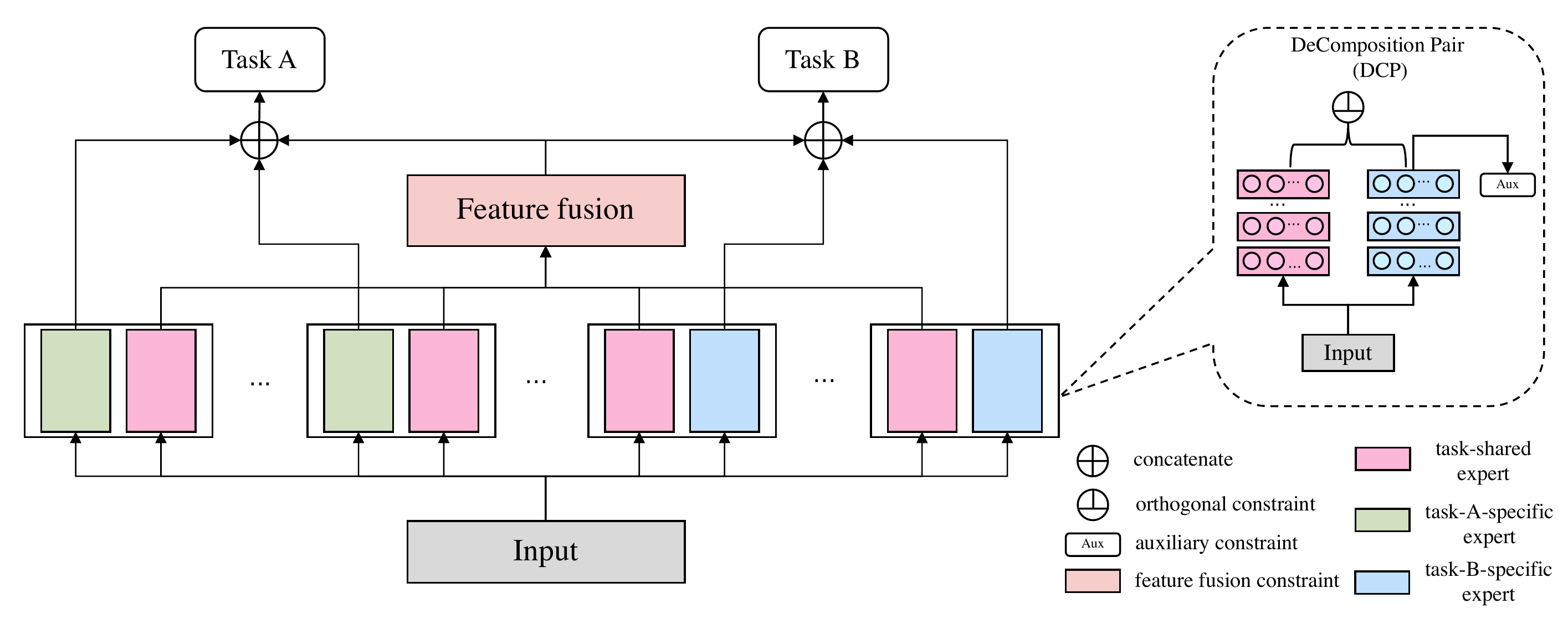}
  \caption{Architecture of the proposed Feature Decomposition Network: 
The gray rectangle box represents the original input, each DeComposition Pair (DCP) consists of two expert networks, including a task-shared expert network (pink box) and task-specific expert network (green or blue box). The detail of the DCPs is shown in the dashed box. An auxiliary task head is added for each DCP to promote feature decomposition.}
  \label{fig:architecture}
\end{figure*}

As shown in Fig.~\ref{fig:architecture}, FDN consists of multiple newly-proposed decomposition pairs (DCPs). For each task, there are $M$ DCP modules. Each DCP module has a pair of experts, a task-specific expert and a task-shared expert. Task-specific expert, which is denoted by $f^{p}$, captures features which are specific for each individual task. Task-shared expert is denoted by $f^{s}$, which extracts features that are potentially shared by all the tasks. After extracting the decomposed features, the model then fuses these features and feeds the features into prediction layer of each task. The process can be represented as follows:

\begin{equation}
f^{MTL}_{k}(X)=\sigma(g_{k}(f^{p}(X),f^{s}(X)))
\label{eq:fdn}
\end{equation}
where $g_{k}(.)$ is a fusion function for task $k$ and $\sigma$ is an activation function.
Next, we will explain these constraints applied to obtain the decomposed features.

\subsection{DeComposition Pair}

As explained before, negative transfer is caused by feature redundancy. To address this problem, our idea is to reduce such redundancy by \textit{explicitly} decomposing features into task-specific and task-shared features. Such design will make the model recombine features based on the requirements of each task. Three constraints are introduced into DCP module.


\subsubsection{Orthogonal Constraint}
Orthogonal constraint has been widely explored for feature decomposition \cite{wan2019generative,salzmann2010factorized,salzmann2010factorized,bousmalis2016domain}. In order to decompose the original features into task-specific features and shared features as much as possible, we add the orthogonal constraint between the extracted features of the two experts. The orthogonal constraint is represented as follows:

\begin{equation}
L_{\text {orth}}=\sum_{k=1}^{K} \sum_{m=1}^{M}\left\|\left(f_{m}^{s}\right)^{\top} f_{m}^{p}\right\|_{F}^{2}
\end{equation}
where $M$ denotes the number of modules of task $k$, $f_{m}^{s}$ denotes the feature extracted by the $m$-th task-shared expert of tash-$k$, and $f_{m}^{p}$ is from the $m$-th task-specific expert of tash-$k$. $\|\cdot\|_{F}^{2}$ denotes squared Frobenius norm.

\subsubsection{Auxiliary Task Constraint}
Orthogonal constraint can help experts capture complementary features, but not necessarily decompose the features into task-shared and task-specific. Therefore, in order to capture task-specific features, we introduce an auxiliary task to each task-specific expert as a regularizer. Specifically, we treat each task-specific expert as a single-task network and directly use its extracted feature for the prediction of a specific task (as shown in  Fig.~\ref{fig:architecture}(Right) rounded rectangular box marked with "aux" text). Such design can induce the task-specific expert to extract the most salient task-specific feature for the current task. It is worth noting that the auxiliary task objective is the same as the final loss of the task. Thus, the corresponding labels are the same. It can be represented as follows:


\begin{equation}
L_{aux}=\sum_{k=1}^{K} \sum_{m=1}^{M} L_{k,m}\left(\hat{y}_{m}^{k}, y^{k}\right)  
\end{equation}

\begin{equation}
\hat{y}_{m}^{k}=\sigma(f_{m}^{k,p}(X))
\end{equation}
where $L_{k,m}(\cdot)$ is the $k$-th auxiliary task loss, and $\hat{y}_{m}^{k}$ denotes the prediction result of using the extracted feature of the $m$-th task-specific expert.

\subsubsection{Task-shared Feature Fusion Constraint}
These task-shared experts of different DCPs can be treated as an fusion model which is able to extract more expressive features than the model using the shared parameters. The fusion of task-shared experts in DCPs is equivalent to a constraint, which will induce the model to further learn task-shared features. Once obtaining the decomposed features, the model fuses these features by concatenation ($g_{k}(.)$ in Eq.~\ref{eq:fdn}), as shown in Fig.~\ref{fig:architecture}.




\subsubsection{Total Loss and Implementation Details}

The final loss function of our model is defined as $L=L_{\text {task }}+L_{\text {orth }}+L_{\text {aux }}$.
For fair comparison, in our implementation, the expert of DCP is designed to be feed-forward network the same as that of MMOE. In addition, in the architecture of MMoE, CGC, and PLE, the design of the tower network is mainly to eliminate the influence of noise on the model performance. In our proposed model, we did not use the tower network as used in MMoE, CGC, PLE, which reduces the amount of parameters, and, as explained before, DCPs have solved the problem of feature redundancy at the expert level and the feature fusion module (as shown in Fig.~\ref{fig:architecture}) has been used to eliminate the noise in advance, thus it is unnecessary to use additional tower networks.

\section{Experiments on Synthetic Dataset}

In the real dataset, due to the complexity of business and the influence of data noise, it is difficult to ideally decompose the original feature input into task specific and task shared features. Therefore, in order to establish the ability of feature decomposition of the models, we use a synthetic data that is easy to verify the decoupling ability of the model.


\subsection{Synthetic Dataset Generation}
Inspired by the previous synthetic data in MMOE \cite{ma2018modeling}, we generated two regression tasks (task A and task B) and generated the ideal synthetic features corresponding to two tasks. In addition, we also generated features for two tasks individually to approximate the ideal performance of the tasks by the single task model. Finally, we generated three datasets, namely Dataset-$A$ : [[task A], [data synthesized by shared features and task A specific features]], Dataset-$B$ : [[task B], [data synthesized by shared features and Task B specific features]], and Dataset-$I$ : [[task A, Task B], [data synthesized by shared features, task A specific features and Task B specific features]], where Dataset-$A$ and Dataset-$B$ are to approach the ideal performance of the two tasks by the single task model respectively, and Dataset-$I$ is the ideal data set finally synthesized to verify the feature decoupling ability of MTL. Further details of generating synthetic data are provided in Appendix.

Since we have generated the dataset of shared features and the specific features of tasks A and B, it is easy to approximate the ideal performance of the tasks. We train the MTL models in the comprehensive synthetic dataset to compare the gap between the ideal performance and results of these MTL models.

\subsection{Performance on Synthetic Data}
We have repeated experiments in the synthetic data for the proposed FDN and the current SOTA multi-task model, and the model hyper-parameters are same in all experiments. We set the embedding size and the number of experts to be 32 and 8 (FDN has 4 pairs of DCPs), the maximum epoch is 5. Each expert network consists of two fully-connected layers (128 units and 64 units).

Mean Square Error (MSE) is used as the evaluation metric.
Table~\ref{tab:idealexp} compares the gap between results achieved by each model and oracle results.
It is clear to see that our proposed FDN model significantly outperforms other baseline methods.

\begin{table}[t]
  \caption{Performance Gap on Synthetic Dataset}
  \centering
  \label{tab:idealexp}
  \resizebox{0.9\columnwidth}{!}{
  \begin{tabular}{ccccc}
    \hline  Model  & Task A/GAP & Task B/GAP \\
    \hline 
    Oracle Results & 0.0\% & 0.0\% \\
    MMoE & -0.2896\% & -0.3166\% \\
    CGC & -0.0580\% & -0.1981\% \\
    PLE & -0.0433\% & -0.0695\% \\
    FDN (ours) & \textbf{-0.0203\%} & \textbf{-0.0367\%} \\
    \hline
      \end{tabular}
  }
  \vspace{-0.5cm}
\end{table}

\section{Experiments on Ali-CCP Dataset}

To demonstrate the effectiveness of the newly proposed FDN, we conducted experiments on ``Alibaba Click and Conversion Prediction'' (Ali-CCP). We take MMoE \cite{ma2018modeling} as the baseline model, and also compare FDN with the SOTA models, CGC and PLE \cite{tang2020progressive}. 




\subsection{Experimental Setup}
Ali-CCP is a dataset gathered from real-world traffic logs of the recommender system in Taobao.  This dataset contains click and conversion data. Therefore, it can be used to train MTL models and evaluated by CTR and CVR. In this dataset, the training dataset and the testing dataset have 42 million samples respectively. We sampled the two subsets without replacement, and extracted 40 million data respectively as the experimental sample set. 

In the experiment, MMoE and CGC are set with 8 experts, and PLE is set with 2 layers and 16 experts. Our model is set with 2 DCPs for each task.

\subsection{Experimental Result}

\begin{figure}[t]
  \centering
  \includegraphics[width=1.0\linewidth]{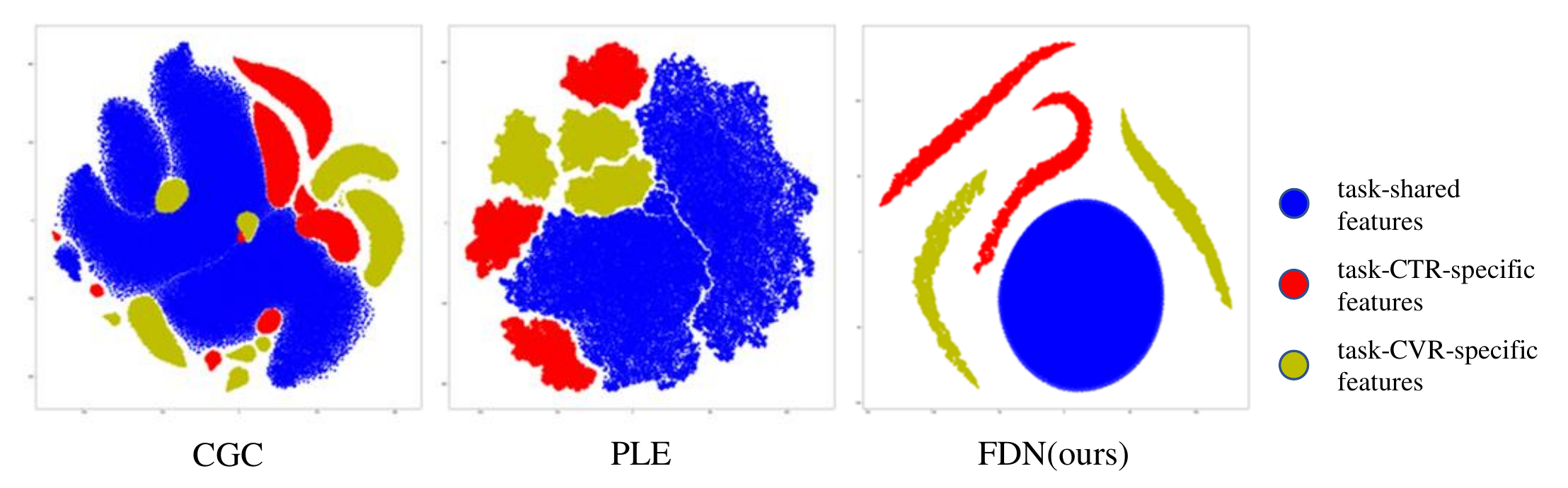}
  \caption{2D space t-SNE feature distributions learned by task-shared experts and task-specific experts of our proposed model in Ali-CCP Dataset.
  }
  \label{fig:aliccp-tsne}
\end{figure}

The experimental results are shown in Table~\ref{tab:freq1}. FDN outperforms other models on both CTR and CVR. We also provide the performance of single-task models for comparison. From the results, we can have the following observations: (1) our proposed model achieves the best or second best performance on both tasks; (2) The performance of the MTL model, MMoE, is lower than single-task models, indicating the hard-sharing architecture can limit the model's expressivity; (3) Our model is efficient since it has the least number of parameters. And in order to show how FDN can better decouple task-specific features and task-shared features, We project the features of experts into 2D space and draw the t-SNE\cite{2008Visualizing} diagram (a non-linear dimensionality reduction technique for embedding high-dimensional data into 2D or 3D space for visualization), Fig.~\ref{fig:aliccp-tsne} shows that FDN distinguishes task-specific features and task-shared features more clearly than other SOTA models.

\begin{table}[t]
  \caption{Quantitative Results on Ali-CPP Dataset}
  \centering
  \label{tab:freq1}
  \resizebox{0.9\columnwidth}{!}{
  \begin{tabular}{cccccc}
    \hline  Model  & CTR/AUC & CVR/AUC &  \#params \\
    \hline 
    Single-Task & 0.6189 & 0.6248 & -\\
    MMoE (baseline) & 0.6175 & 0.6239 & $5.75\times10^{9}$\\
    CGC & 0.6235 & 0.6616 & $6.39\times10^{9}$\\
    PLE & 0.6241 & 0.6308 & $7.03\times10^{9}$\\
    FDN (ours) & \textbf{0.6252} & \textbf{0.6783} & $5.11\times10^{9}$\\
    \hline
  \end{tabular}
  }
\end{table}



\begin{table}[t]
  \caption{Ablation Study on Ali-CPP Dataset}
  \centering
  \label{tab:ablation}
  \resizebox{0.9\columnwidth}{!}{
  \begin{tabular}{ccccc}
    \hline  Model  & CTR/AUC & CVR/AUC \\
    \hline 
    FDN(without aux) & 0.6151 & 0.6610 \\
    FDN(without orth) & 0.6193 & 0.6615 \\
    FDN(without task-shared feature) & 0.6143 & 0.6558 \\
    FDN & \textbf{0.6252} & \textbf{0.6783} \\
    \hline
      \end{tabular}
  }
\end{table}

\subsection{Abalation Study} We further conducted several experiments to show the contribution of each component in our model. As shown in Table~\ref{tab:ablation}, we remove each component from the full model, including the orthogonal constraint (FDN(w/o orth.)), auxiliary task (FDN(w/o aux.)) and task-shared feature fusion constraint (FDN(w/o task-shared feature.)). 
It is clearly to see that all the component contributes to the final performance to a certain degree.

\section{Conclusion}
In this paper, we have proposed a novel multi-task learning method, called FDN, for recommender system. The core of the FDN is a new feature decomposition module with two carefully-designed constraints. Besides, an ensemble model is adopted to further improve the expressivity of the extracted feature.
We have conducted experiments on two datasets and achieved new SOTA performance.

\section*{Acknowledgements}
This work was supported by the National Natural Science Foundation of China (No. 62202025 and 62002012).

\bibliography{aaai23}

\appendix
\section{Appendix}
\subsection{Synthetic Data Generation}
Synthetic data generation steps as follow:
\begin{itemize}
\item [1)]
Randomly generated unit vector $u_{1}, u_{2}, u_{s} \in R^{d}$ :
\begin{equation}
\left\|u_{1}\right\|_{2}=1,\left\|u_{2}\right\|_{2}=1,\left\|u_{s}\right\|_{2}=1
\end{equation}

\item [2)]
Given constant $c_{1}, c_{2}, c_{s}$ generate weight vector:
\begin{equation}
w_{1}=c_{1} u_{1}, w_{2}=c_{2} u_{2}, w_{s}=c_{s} u_{s}
\end{equation}

\item [3)]
Randomly generated $x_{1}, x_{2}, x_{s} \in R^{d}$, where $x_{1}^{i} \sim N(-1,0.1), x_{2}^{i} \sim N(1,0.1), x_{s}^{i} \sim N(0,1)$.

\item [4)]
Generate two labels for regression tasks:
\begin{equation}
y_{1}=s+w_{1}^{T} x_{1}+\sum_{i=1}^{m} \sin \left(\alpha_{i} w_{1}^{T} x_{1}+\beta_{i}\right)+\epsilon_{1}
\end{equation}
\begin{equation}
y_{2}=s+w_{2}^{T} x_{2}+\sum_{i=1}^{m} \sin \left(\alpha_{i} w_{2}^{T} x_{2}+\beta_{i}\right)+\epsilon_{2}
\end{equation}
where $s=w_{s}^{T} x_{s}+\sum_{i=1}^{m} \sin \left(\alpha_{i} w_{s}^{T} x_{s}+\beta_{i}\right), \epsilon_{1}, \epsilon_{2} \stackrel{\text { i.i.d }}{\sim} N(0,0.01)$

\item [5)]
Generate features:
\begin{equation}
\begin{aligned}
X_{1 i}=& \sin \left(\delta_{i}\left(u_{1}^{i} x_{1}^{i}+u_{s}^{i} x_{s}^{i}\right)+\gamma_{i}\right) + \\ & \cos \left(\delta_{i}\left(u_{1}^{i} x_{1}^{i}+u_{s}^{i} x_{s}^{i}\right)+\gamma_{i}\right)+\varepsilon
\end{aligned}
\end{equation}
\begin{equation}
\begin{aligned}
X_{2 i}=& \sin \left(\delta_{i}\left(u_{2}^{i} x_{2}^{i}+u_{s}^{i} x_{s}^{i}\right)+\gamma_{i}\right)+ \\ & \cos \left(\delta_{i}\left(u_{2}^{i} x_{2}^{i}+u_{s}^{i} x_{s}^{i}\right)+\gamma_{i}\right)+\varepsilon
\end{aligned}
\end{equation}
\begin{equation}
\begin{aligned}
X_{i}=&\sin \left(\delta_{i}\left(u_{1}^{i} x_{1}^{i}+u_{2}^{i} x_{2}^{i}+u_{s}^{i} x_{s}^{i}\right)+\gamma_{i}\right)+ \\ & \cos \left(\delta_{i}\left(u_{1}^{i} x_{1}^{i}+u_{2}^{i} x_{2}^{i}+u_{s}^{i} x_{s}^{i}\right)+\gamma_{i}\right)+\varepsilon
\end{aligned}
\end{equation}
where $\varepsilon \stackrel{\text { i.i.d }}{\sim} N(0,0.01)$ , $X_{1 i}$ and $X_{2 i}$ as a feature of the single task model, $X_{i}$ as a feature of multitasking dataset.

\item [6)]
Repeat 3), 4) and 5) until enough data are generated.
\end{itemize}

\balance

\end{document}